\documentclass[reprint,amsmath,amssymb,aps]{revtex4-2}
\usepackage{graphicx}
\usepackage{bm}
\usepackage{mathrsfs}
\usepackage{slashed}
\usepackage{makecell}
\allowdisplaybreaks[4]
\usepackage[colorlinks,citecolor=blue,urlcolor=blue,linkcolor=blue]{hyperref}

\begin{document}

    \preprint{APS/123-QED}

	\title{Semileptonic Decay of $\Xi_c \to \Xi \ell^+ \nu_\ell$ From Light-Cone QCD Sum Rules}

    \author{Hui-Hui Duan}
    \email{duanhuihui19@nudt.edu.cn}
    \author{Yong-Lu Liu}
    \email{yongluliu@nudt.edu.cn}
    \author{Ming-Qiu Huang}
    \email{mqhuang@nudt.edu.cn (corresponding author)}

	\affiliation{Department of Physics, National University of Defense Technology, Changsha 410073, Hunan, People's Republic of China}

	\begin{abstract}
     Semileptonic decay processes of $\Xi_c \to \Xi \ell \nu_\ell$ are studied by light-cone QCD sum rules in this paper. The six form factors of $\Xi_c \to \Xi$ semileptonic transition matrix elements are calculated by this method with the light-cone distribution amplitudes of $\Xi$ baryon up to twist six. With the six form factors, the absolute branching ratios of $\Xi_c^0 \to \Xi^- \ell^+ \nu_\ell$ and $\Xi_c^+ \to \Xi^0 \ell^+ \nu_\ell$ are calculated by the helicity amplitudes formalism of semileptonic differential decay widths. The ratios of absolute branching ratios of electron and muon final-states processes give the proof of lepton flavor universality. Our results are in accordance with the recent experimental and theoretical reports.
	\end{abstract}
     \maketitle
     
	\section{\label{sec:I}Introduction}  
     
      Recently, ALICE and Belle reported the measurement of decay branching ratios of $\Xi_c^0 \to \Xi^-e^+\nu_e$ semileptonic decays, respectively. ALICE gave the relative branching ratio $\Xi_c^0 \to \Xi^-e^+\nu_e$ to $\Xi_c^0 \to \Xi^- \pi^+$ result $\Gamma(\Xi_c^0 \to \Xi^- e^+ \nu_e)/\Gamma(\Xi_c^0 \to \Xi^- \pi^+)=1.38\pm 0.14(stat)\pm0.22(syst)$ \cite{Zhu:2020yci, ALICE:2021bli}. The absolute branching ratio of $\Xi_c^0 \to \Xi^- \pi^+$ was first measured by Belle in 2019, with the suggestion value $(1.8\pm0.50\pm0.14)\%$ \cite{Belle:2018kzz}. And, the first measurement of the absolute branching ratio of $\Xi_c^+ \to \Xi^-\pi^+\pi^+$ was also given by Belle in 2019 \cite{Belle:2019bgi}. With the combination of the relative branching ratio ALICE measured and the absolute branching ratio of $\Xi_c^0 \to \Xi^-\pi^+$ which Belle measured, the ALICE Collaboration's result of decay absolute branching ratio of $\Xi_c^0 \to \Xi^- e^+ \nu_e$ is around $2.48 \%$.  The Belle Collaboration reported the branching ratios of positron final state $\mathcal{B}(\Xi_c^0 \to \Xi^- e^+ \nu_e)=(1.31\pm0.04\pm0.07\pm0.38)\%$ and muon final state $\mathcal{B}(\Xi_c^0 \to \Xi^- \mu^+ \nu_\mu)=(1.27\pm0.06\pm0.10\pm0.37)\%$ \cite{Belle:2021crz}, and the third uncertainty above comes from the measurement of absolute branching ratio of $\Xi_c^0 \to\Xi^-\pi^+$, while the first and second uncertainties are statistical and theoretical. In the early experiments, ARGUS and CLEO Collaborations gave their measurement of relative branching ratio of $\mathcal{B}(\Xi_c^0\to\Xi^-e^+\nu_e)/\mathcal{B}(\Xi_c^0\to\Xi^-\pi^+)=0.96\pm0.43\pm0.18$ and $3.1\pm1.0^{+0.3}_{-0.5}$ in 1990s, respectively \cite{ARGUS:1992jnv, CLEO:1994aud}; and a review in 1990s see Ref. \cite{Richman:1995wm}. Therefore, in the experiment aspect, the semileptonic decay results of $\Xi_c$ baryons have a wide range. It is worth investigating these processes in theoretical and experimental aspects by further studies.
      
      With the phenomenology models, some articles have investigated these processes, and still a wide range in these analyses. In the first lattice calculation recently \cite{Zhang:2021oja}, it indicated that the semileptonic decay branching ratios of $\Xi_c \to \Xi$ are 
      \begin{align*}
      	\mathcal{B}(\Xi_c^0 \to \Xi^- e^+ \nu_e)=2.38(0.30)_{stat.} (0.32)_{syst.} \%, \notag \\
      	\mathcal{B}(\Xi_c^0 \to \Xi^- \mu^+ \nu_\mu)=2.29(0.29)_{stat.} (0.31)_{syst.} \%, \notag \\
      	\mathcal{B}(\Xi_c^+ \to \Xi^0 e^+ \nu_e)=7.18(0.90)_{stat.} (0.98)_{syst.} \%, \notag \\
      	\mathcal{B}(\Xi_c^+ \to \Xi^0 \mu^+ \nu_\mu)=6.91(0.87)_{stat.} (0.93)_{syst.} \%. \notag \\
      \end{align*}
      This is in accordance with the ALICE results but larger than the Belle.  Many other phenomenology models such as the light-cone QCD sum rule method \cite{Liu:2009uc,Liu:2010bh,Azizi:2011mw,Aliev:2021wat}, SU(3) flavor symmetry model \cite{Geng:2018plk, Geng:2019bfz}, light-front quark model \cite{Zhao:2018zcb,Geng:2020gjh,Ke:2021pxk}, relativistic quark model \cite{Faustov:2019ddj} and QCD sum rules \cite{Zhao:2021sje} {\it et al.} give the branching ratio of $\Xi_c^0 \to \Xi^- \ell^+ \nu_\ell$ from $1.35\%$ to $(7.26\pm2.54)\%$, and $\Xi_c^+ \to \Xi^0 \ell^+ \nu_\ell$ from $3.38^{+2.19}_{-2.26}\%$ to $(11.9\pm1.3)\%$. These models also list a wide range of these processes, so it still leads to a more charming calculation to pursue.
      
      In this article, we study the $\Xi_c \to \Xi$ semileptonic decay with the light-cone QCD sum rules. In our previous works, the light-cone distribution amplitudes of octet baryon $\Xi$ have been obtained, and the related parameters of $\Xi_c$ and $\Xi$ obtained from QCD sum rules too \cite{Liu:2009uc,Liu:2010bh}. With the $\Xi$ baryon light-cone distribution amplitudes, decuplet baryon $\Omega_c^0 \to \Xi^- \ell^+\nu_\ell$ decay branching ratio have also given in the same method \cite{Duan:2020xcc}. And other usage on the calculation of the electromagnetic nucleon form factors can be found in \cite{Anikin:2013aka}. Calculation of the form factors of the weak decay transition matrix element of heavy baryon to light baryon makes known the information of heavy baryon properties, especially the decay branching ratios. Based on these, the weak semileptonic decays of $\Xi_c \to \Xi \ell^+ \nu_\ell$ are calculated in this work by the method and parameters mentioned above.
      
     The theoretical framework of this work and the formalism results are presented in Sec. \ref{sec:II}, and the six form factors of $\Xi_c$ to $\Xi$ transition are calculated by light-cone QCD sum rules. Numerical analyses of the semileptonic processes are given in Sec. \ref{sec:III}. The conclusions are given in Sec. \ref{sec:IV}.
      
	\section{\label{sec:II}Light-cone sum rules of $\Xi_c \to \Xi$ transition} 

	Light-cone QCD sum rules are developed from the standard QCD sum rule proposed by Shifman, Vainshtein and Zakharov, which is often called SVZ QCD sum rules \cite{Shifman:1978bx} and extended to light-cone by Chernyak {\it et al.} \cite{Chernyak:1983ej}. They all start with the hadron correlation function, then parameterized by hadron states on the hadronic level and expanded it at QCD level by operator product expansion. The difference is that the conventional SVZ QCD sum rules make the operator product expansion at position point $x=0$ with mass dimensions and light-cone QCD sum rules make the operator product expansion at light-cone $(x^2=0)$ with increasing twists (dimension minus spin). The basic nonperturbative input parameters in conventional SVZ QCD sum rules are vacuum condensate, while in light-cone QCD sum rules they are parametrized to light-cone distribution amplitudes \cite{Bakulev:2001pa,Mikhailov:2010ud}. The application of light-cone QCD sum rules to semileptonic decay of heavy baryon transition to light baryon with light-cone distribution amplitudes have two versions: one is using the final heavy-baryon state light-cone distribution amplitudes and another is using the initial heavy baryon's. Light-cone QCD sum rules were first used in $\Lambda_b\to p\ell\overline{\nu}$ with light baryon distribution amplitude in Ref. \cite{Huang:2004vf}, and with the heavy-baryon version was first studied in the processes $\Lambda_b \to p, \Lambda$ semileptonic decay in Ref. \cite{Wang:2009hra}. The light-baryon distribution amplitudes version is adopted in the following calculation.
	
	To calculate the decay properties of heavy baryons one needs to know the decay matrix element. Matrix element of heavy baryon decay to light baryon $\Xi_c \to \Xi$ can be parameterized as six form factors by the following form, 
    \begin{widetext}
	\begin{align}
		\langle \Xi_c^{(*)}(P')|j_\nu|\Xi(p)\rangle=&\bar{u}_{\Xi_c^{(*)}}(P')\{f_1^{(*)}(q^2)\gamma_\nu+ i \frac{f_2^{(*)}(q^2)}{M_{\Xi_c^{(*)}}} \sigma_{\nu\mu}q^\mu +  \frac{f_3^{(*)}(q^2)}{M_{\Xi_c^{(*)}}}q_\nu \notag \\&-[g_1^{(*)}(q^2)\gamma_\nu+i \frac{g_2^{(*)}(q^2)}{M_{\Xi_c^{(*)}}} \sigma_{\nu\mu} q^\mu +\frac{g_3^{(*)}(q^2)}{M_{\Xi_c^{(*)}}}q_\nu]\gamma_5\}u_{\Xi}(p), \label{transition element}
	\end{align}
    \end{widetext}	
	where $f_i(q^2)/g_i(q^2))(i=1,2,3)$ are the weak decay form factors, $M_{\Xi_c}$ is the mass of $\Xi_c$ baryon, $P'=p-q$, $p$ is the momentum of $\Xi$, $q$ is the momentum transfer and $u_{\Xi_c}$ and $u_\Xi$ are the spinors of $\Xi_c$ and $\Xi$, respectively. It is known that the negative parity of interpolating baryon makes contributions in the QCD sum rules \cite{Jido:1996ia,Khodjamirian:2011jp,Pimikov:2019dyr}. So, the asterisks stand for the form factors, spinor and mass of $\Xi_c^*$ baryon with spin-parity $\frac{1}{2}^-$  transition.
	
	In order to obtain the light-cone sum rules of these form factors, one begins with the two-point correlation function sandwiched between vacuum and final $\Xi$ baryon state:
	\begin{gather}
		T_\nu (p,q)=i\int d^4x e^{iq \cdot x}\langle 0|T\{j_{\Xi_c}(0)j_\nu (x)\}|\Xi(p)\rangle  \label{correlation function}
	\end{gather}
	where the $j_{\Xi_c}(0)$ and $j_\nu (x)$ are heavy-baryon $\Xi_c$ current and weak decay current respectively. In this study, the Ioffe-type current is chosen on the quark level for this computation:
	\begin{gather}
		j_{\Xi_c}(x)=\epsilon_{ijk}[s^{iT}(x)C\gamma_\mu c^j(x)]\gamma_5\gamma^\mu q^k(x),   \label{baryon current}
	\end{gather} 
	and the weak decay current is
	\begin{gather}
		j_\nu(x)=\bar{c}(x)\gamma_\nu(1-\gamma_5)s(x).    \label{weak current}
	\end{gather}

  With the quark-hadron duality, one can parameterize the correlation function on both hadronic and QCD level. Using the completness relation of $\Xi_c$
    \begin{gather}
    	\int d^4P' \sum_i|\Xi_c^i(P')\rangle \langle \Xi_c^i(P')|=1,
    \end{gather}
  where the index $i$ contains all states with the quantum number of $\Xi_c$. By using the dispersion relation, one can separate the lowest states of hadron $\Xi_c(\Xi_c^*)$, and the correlation can be expressed on the hadronic level by
  \begin{align}
  	T_\nu(p,q)=&\frac{\langle 0|j_{\Xi_c}|\Xi_c(P')\rangle \langle \Xi_c(P')|j_\nu |\Xi(p)\rangle}{M_{\Xi_c}^2-P'^2} \notag \\& +\frac{\langle 0|j_{\Xi_c^*}|\Xi_c^*(P')\rangle \langle \Xi_c^*(P')|j_\nu |\Xi(p)\rangle}{M_{\Xi_c^*}^2-P'^2}+\cdots,
  \end{align}
  the ellipsis contains the contribution of high resonance and continuum states. The baryon to vacuum transition matrix element is defined by
  \begin{align}
  	\langle 0|j_{\Xi_c}|\Xi_c(P')\rangle=f_{\Xi_c}u_{\Xi_c}(P'),
  	\end{align}
  	and 
  	\begin{align}
  		\langle 0|j_{\Xi_c^*}|\Xi_c^*(P')\rangle=f_{\Xi_c^*}\gamma_5u_{\Xi_c^*}(P'),
   \end{align} 
    where$f_{\Xi_c}$ ($f_{\Xi_c^*}$) is the decay constant of $\Xi$ ($\Xi_c^*$) baryon. The decay constant $f_{\Xi_c}$ has been calculated in Ref.~\cite{Liu:2009uc,Liu:2010bh,Azizi:2011mw,Wang:2010fq,Shi:2019hbf}; we will adopt the value $f_{\Xi_c}=0.038 \rm{GeV^3}$ which is given by~\cite{Shi:2019hbf,Shi:2022kfa} in the following calculations. The decay constant of $f_{\Xi_c^*}$ does not contribute in the final sum rules. With the ralations of baryon Dirac spinors $u_{\Xi_c^{(*)}}(P',s)$ and summing up the spin 
    \begin{equation}
    \sum_s u_{\Xi_c^{(*)}}(P',s)\overline{u}_{\Xi_c^{(*)}}(P',s)=\slashed{P}'+M_{\Xi_c^{(*)}},
    \end{equation}
    the hadronic representation of this transition process at the hadronic level can be expressed as
   \begin{widetext}
    \begin{align}
    	T_\nu (P',q^2)=&\frac{f_{\Xi_c}}{M_{\Xi_c}^2-P'^2}\{[(M_{\Xi_c}-M_\Xi)f_1(q^2)+\frac{2p\cdot q-q^2}{M_{\Xi_c}}f_2(q^2)]\gamma_\nu+[\frac{M_{\Xi_c}+M_\Xi}{M_{\Xi_c}}(f_2(q^2)+f_3(q^2))-2f_1(q^2)]q_\nu \notag \\& +(f_1(q^2)-\frac{M_{\Xi_c}+M_\Xi}{M_{\Xi_c}}f_2(q^2))\gamma_\nu \slashed{q}+2f_1(q^2) p_\nu -\frac{2f_2(q^2)}{M_{\Xi_c}}p_\nu \slashed{q}+\frac{f_2(q^2)-f_3(q^2)}{M_{\Xi_c}}q_\nu \slashed{q} \notag \\ & -[(M_{\Xi_c}+M_\Xi)g_1(q^2)+\frac{2p\cdot q-q^2}{M_{\Xi_c}}g_2(q^2)]\gamma_\nu\gamma_5+[\frac{M_{\Xi_c}-M_\Xi}{M_{\Xi_c}}(g_2(q^2)+g_3(q^2))+2g_1(q^2)]q_\nu \gamma_5 \notag \\ & +(\frac{M_{\Xi_c}-M_\Xi}{M_{\Xi_c}}g_2(q^2)-g_1(q^2))\gamma_\nu \slashed{q}\gamma_5-2g_1(q^2) p_\nu\gamma_5 +\frac{2g_2(q^2)}{M_{\Xi_c}}p_\nu \slashed{q}\gamma_5+\frac{g_3(q^2)-g_2(q^2)}{M_{\Xi_c}}q_\nu \slashed{q}\gamma_5\}u_{\Xi}(p) \notag \\& +\frac{f_{\Xi_c^*}}{M_{\Xi_c^*}^2-P'^2}\{[(M_\Xi-M_{\Xi_c^*})f_1^*(q^2)-\frac{2p\cdot q-q^2}{M_{\Xi_c^*}}f_2^*(q^2)]\gamma_\nu\gamma_5+[\frac{M_{\Xi_c^*}+M_\Xi}{M_{\Xi_c^*}}(f_2^*(q^2)+f_3^*(q^2)) \notag \\& -2f_1^*(q^2)]q_\nu\gamma_5-(\frac{M_{\Xi_c^*}+M_\Xi}{M_{\Xi_c^*}}f_2^*(q^2)-f_1^*(q^2))\gamma_\nu\slashed{q}\gamma_5+2f_1^*(q^2)p_\nu\gamma_5+\frac{2f_2^*(q^2)}{M_{\Xi_c^*}}p_\nu\slashed{q}\gamma_5 \notag \\&-\frac{f_2^*(q^2)-f_3^*(q^2)}{M_{\Xi_c^*}}q_\nu\slashed{q}\gamma_5+[(M_{\Xi_c^*}+M_\Xi)g_1^*(q^2)+\frac{2p\cdot q-q^2}{M_{\Xi_c^*}}g_2^*(q^2)]\gamma_\nu+[\frac{M_\Xi-M_{\Xi_c^*}}{M_{\Xi_c^*}}(g_2^*(q^2) \notag \\& +g_3^*(q^2)) +2g_1^*(q^2)]q_\nu-(\frac{M_\Xi-M_{\Xi_c^*}}{M_{\Xi_c^*}}g_2^*(q^2)+g_1^*(q^2))\gamma_\nu\slashed{q}-2g_1^*(q^2)p_\nu-\frac{2g_2^*(q^2)}{M_{\Xi_c^*}}p_\nu\slashed{q} \notag \\& +\frac{g_2^*(q^2)-g_3^*(q^2)}{M_{\Xi_c^*}}q_\nu\slashed{q}\}u_{\Xi}(p)+\cdots.
    \end{align}
    \end{widetext}
    Where $M_\Xi$ is the mass of $\Xi$ baryon. 
   
    On the theoretical side, the correlation function will be calculated by contracting the heavy charm quark, and be derived on the quark level with current quark model. The correlation function can be written as
   \begin{widetext}  
   \begin{align}
   	T_\nu (p,q)=-i\int d^4 x e^{iq\cdot x}[C\gamma_\mu S(-x)\gamma_\nu(1-\gamma_5)]_{\alpha\tau}(\gamma_5\gamma^\mu)_{\sigma\gamma}\langle 0|\epsilon^{ijk}s^i_\alpha (0)s^j_\tau (x)u^k_\gamma (0)|u_\Xi (p)\rangle,
   \end{align}
   \end{widetext}
    where $C$ is charge conjugation, and $S(-x)$ is the free charm quark propagator
    \begin{align}
     	S(-x)=i\int d^4 x \frac{e^{ik\cdot x}}{\slashed{k}-m_c}.
    \end{align}
   
   In the calculation of correlation function, the matrix element $\langle 0|\epsilon^{ijk}s_\alpha^i(0)s^j_\lambda(x)q^k_\sigma(0)|u_\Xi(p)\rangle$ can be expressed with the light-cone distribution amplitudes of octet baryon $\Xi$. It has been developed from nucleon distribution amplitudes from leading twist to twist six and the explicit expressions have also been given by our previous works \cite{Chernyak:1987nu,Braun:2000kw,Liu:2008yg,Liu:2009uc}. The standard procedure of light-cone QCD sum rules calculations on the theoretical side gives 12 structures, each of them corresponding to the same structure of the hadronic side. However, as referred to in the previous discussion, due to the negative parity particle contribution in the sum rules, we should tackle the problem of the negative parity issue. For the aim to subtract the negative parity baryon contribution, we solve the linear equations corresponding to the same Lorentz structure to the hadronic side and QCD side as the same method in \cite{Khodjamirian:2011jp}. After that, one obtains the six form factors of $\Xi_c$ to $\Xi$ transition matrix element; there will be without the negative parity parts, and only the negative parity baryon $\Xi_c^*$ mass enter the sum rules.
    
   To suppress higher twists contribution, the Borel transform operation is made for every form factor on both hadronic and QCD side. This introduces a Borel parameter $M_B$ in the form factors; it will be discussed in the numerical analysis section.  After making the Borel transformation, the form factors can be written as
 \begin{widetext}
   \begin{align}
   f_1(q^2)=&\frac{e^{M_{\Xi_c}^2/M_B^2}}{2f_{\Xi_c}(M_{\Xi_c}+M_{\Xi_c^*})}\{(M_\Xi+M_{\Xi_c})[(M_\Xi-M_{\Xi_c})\Pi_{p_\nu \slashed{q}}+\Pi_{p_\nu}]+2(M_{\Xi_c^*}-M_{\Xi_c})\Pi_{\gamma_\nu \slashed{q}}+2\Pi_{\gamma_\nu}\}, \label{f1} \\
   f_2(q^2)=&\frac{M_{\Xi_c}e^{M_{\Xi_c}^2/M_B^2}}{2f_{\Xi_c}(M_{\Xi_c}+M_{\Xi_c^*})}\{M_\Xi \Pi_{p_\nu \slashed{q}}-M_{\Xi_c^*}\Pi_{p_\nu \slashed{q}}+\Pi_{p_\nu}-2\Pi_{\gamma_\nu \slashed{q}}\},   \\
  f_3(q^2)=&\frac{M_{\Xi_c}e^{M_{\Xi_c}^2/M_B^2}}{2f_{\Xi_c}(M_{\Xi_c}+M_{\Xi_c^*})}\{(M_\Xi-M_{\Xi_c^*})(\Pi_{p_\nu \slashed{q}}+2\Pi_{q_\nu \slashed{q}})+\Pi_{p_\nu}+2(\Pi_{\gamma_\nu \slashed{q}}+\Pi_{q_\nu})\},   \\
  g_1(q^2)=&\frac{e^{M_{\Xi_c}^2/M_B^2}}{2f_{\Xi_c}(M_{\Xi_c}+M_{\Xi_c^*})}\{(M_\Xi-M_{\Xi_c^*})[(M_\Xi+M_{\Xi_c^*})\Pi_{p_\nu \slashed{q}\gamma_5}-\Pi_{p_\nu \gamma_5}]+2(M_{\Xi_c}-M_{\Xi_c^*})\Pi_{\gamma_\nu \slashed{q}\gamma_5}-2\Pi_{\gamma_\nu \gamma_5}\},  \\
  g_2(q^2)=&\frac{M_{\Xi_c}e^{M_{\Xi_c}^2/M_B^2}}{2f_{\Xi_c}(M_{\Xi_c}+M_{\Xi_c^*})}\{(M_\Xi+M_{\Xi_c^*})\Pi_{p_\nu \slashed{q}\gamma_5}+2\Pi_{\gamma_\nu \slashed{q}\gamma_5}-\Pi_{p_\nu \gamma_5}\}, \\
  g_3(q^2)=&\frac{M_{\Xi_c}e^{M_{\Xi_c}^2/M_B^2}}{2f_{\Xi_c}(M_{\Xi_c}+M_{\Xi_c^*})}\{(M_\Xi+M_{\Xi_c^*})(2\Pi_{q_\nu\slashed{q}\gamma_5}+\Pi_{p_\nu \slashed{q}\gamma_5})-2\Pi_{\gamma_\nu\slashed{q}\gamma_5}-2\Pi_{q_\nu\gamma_5}-\Pi_{p_\nu\gamma_5}\}. \label{g3}
   	\end{align}
  Where these terms of $\Pi_\Gamma$ in the form factors are the coefficients of the corresponding Lorentz structure $\Gamma=\{p_\nu, \gamma_\nu, p_\nu\slashed{q}, \gamma_\nu\slashed{q}, q_\nu, q_\nu\slashed{q}, p_\nu\gamma_5, \gamma_\nu\gamma_5, p_\nu\slashed{q}\gamma_5, \gamma_\nu\slashed{q}\gamma_5, q_\nu\gamma_5, q_\nu\slashed{q}\gamma_5\}$ on the QCD side, and they have been made the Borel transformation. The general expression of $\Pi_\Gamma$ in our sum rules is given by
  \begin{align}
  	\Pi_\Gamma=&-\int_{\alpha_{20}}^{1}\frac{d\alpha_2}{\alpha_2}\rho_\Gamma^1(\alpha_2)e^{-s/M_B^2} \notag \\&
  	   	+\frac{1}{M_B^2}\int_{\alpha_{20}}^{1}\frac{d\alpha_2}{\alpha_2^2}\rho_\Gamma^2(\alpha_2, q^2)e^{-s/M_B^2}+\frac{\rho_\Gamma^2(\alpha_{20}, q^2)e^{-s_0/M_B^2}}{\alpha_{20}^2M_\Xi^2-q^2+m_c^2} \notag \\&
  	   	-\frac{1}{2M_B^4}\int_{\alpha_{20}}^{1}\frac{d\alpha_2}{\alpha_2^3}\rho_\Gamma^2(\alpha_2, q^2)e^{-s/M_B^2}-\frac{1}{2}\frac{\rho_\Gamma^3(\alpha_{20}, q^2)e^{-s_0/M_B^2}}{\alpha_{20}M_B^2(\alpha_{20}^2M_\Xi^2-q^2+m_c^2)} \notag \\&
   		+\frac{1}{2}\frac{\alpha_{20}^2}{\alpha_{20}^2M_\Xi^2-q^2+m_c^2}[\frac{d}{d\alpha_{20}}\frac{\rho_\Gamma^3(\alpha_{20}, q^2)}{\alpha_{20}(\alpha_{20}^2M_\Xi^2-q^2+m_c^2)}]e^{-s_0/M_B^2}.  \label{sum rule}
   \end{align}
   The detail expressions of $\rho_\Gamma^i (i=1,2,3)$ in the integrands are given in the Appendix.

    In the above form factors, the parameter $s$ on the exponential is 
    \begin{equation}
    s=(1-\alpha_2)M_\Xi^2-\frac{1-\alpha_2}{\alpha_2}q^2+\frac{m_c^2}{\alpha_2}.
    \end{equation}
    The $\alpha_{20}$ which is related to the threshold $s_0$ is expressed as 
    \begin{equation}
    \alpha_{20}=\frac{(q^2-s_0+M_\Xi^2)+\sqrt{(q^2-s_0+M_\Xi^2)-4M_\Xi^2(q^2-m_c^2)}}{2M_\Xi^2}.
    \end{equation}
    $\alpha_i (i=1,2,3)$ are the components of quark coordinate. The relation of $\alpha_i$ satisfy $1=\alpha_1+\alpha_2+\alpha_3$.
    \end{widetext}
    
      \begin{table*}
      	\caption{\label{tab1}Distribution amplitudes with increasing twists to six.} 
      	\begin{ruledtabular}
      		\begin{tabular}{cccc}
      			Twist-3&Twist-4&Twist-5&Twist-6 \\ \hline
      			$\makecell[l]{V_1(x_i)=120x_1 x_2 x_3\phi_3^0,  \\
      				A_1(x_i)=0, \\
      				T_1(x_i)=120x_1 x_2 x_3\phi_3^{'0}.}$&$\makecell[l]{V_2(x_i)=24x_1x_2\phi_4^0,  \\
      				A_2(x_i)=0, \\
      				V_3(x_i)=12x_3(x_1-x_2)\psi_4^0,  \\
      				A_3(x_i)=-12x_3(x_1-x_2)\psi_4^0,  \\
      				T_2(x_i)=24x_1x_2\phi_4^{'0},  \\
      				T_3(x_i)=6x_3(1-x_3)(\xi_4^0+\xi_4^{'0}),  \\
      				T_7(x_i)=6x_3(1-x_3)(\xi_4^{'0}-\xi_4^0).}$&$\makecell[l]{V_4(x_i)=3(x_1-x_3)\psi_5^0, \\
      				A_4(x_i)=3(x_1-x_2)\psi_5^0, \\
      				V_5(x_i)=6x_3\phi_5^0, \\
      				A_5(x_i)=0, \\
      				T_4(x_i)=-\frac{3}{2}(x_1+x_2)(\xi_5^{'0}+\xi_5^0), \\
      				T_5(x_i)=6x_3\phi_5^{'0}, \\
      				T_8(x_i)=\frac{3}{2}(x_1+x_2)(\xi_5^{'0}-\xi_5^0).}$&$\makecell[l]{V_6(x_i)=2\phi_6^0, \\
      				A_6(x_i)=0, \\
      				T_6(x_i)=2\phi_6^{'0}.}$ \\ 
      		\end{tabular}
      	\end{ruledtabular}
      \end{table*}
    
    The light-cone distribution amplitudes used above are defined as follows:
    \begin{align}
	D_0(\alpha_2)=&\int_{0}^{1-\alpha_2}d\alpha_1 V_1(\alpha), \notag \\
	D_1(\alpha_2)=&\int_{0}^{\alpha_2}d\alpha_2'\int_{0}^{1-\alpha_2'}(V_1-V_2-V_3)(\alpha '), \notag \\
	D_2(\alpha_2)=&\int_{0}^{1-\alpha_2}d\alpha_1 V_3(\alpha), \notag \\
	D_3(\alpha_2)=&\int_{0}^{\alpha_2}d\alpha_2'\int_{0}^{1-\alpha_2'}d\alpha_1(-2V_1+V_3+V_4 +2V_5)(\alpha '), \notag \\
	D_4(\alpha_2)=&\int_{0}^{\alpha_2}d\alpha_2'\int_{0}^{1-\alpha_2'}d\alpha_1(V_4-V_3)(\alpha '), \notag \\
	D_5(\alpha_2)=&\int_{0}^{\alpha_2}d\alpha_2'\int_{0}^{\alpha_2'}d\alpha_2''\int_{0}^{1-\alpha_2''}d\alpha_1(-V_1+V_2 \notag \\& +V_3+V_4+V_5-V_6)(\alpha ''), \notag \\
	E_0(\alpha_2)=&\int_{0}^{1-\alpha_2}d\alpha_1A_1(\alpha), \notag \\
	E_1(\alpha_2)=&\int_{0}^{\alpha_2}d\alpha_2'\int_{0}^{1-\alpha_2'}d\alpha_1(-A_1+A_2-A_3)(\alpha '), \notag \\
	E_2(\alpha_2)=&\int_{0}^{1-\alpha_2}d\alpha_1A_3(\alpha), \notag \\
	E_3(\alpha_2)=&\int_{0}^{\alpha_2}d\alpha_2'\int_{0}^{1-\alpha_2'}d\alpha_1(-2A_1-A_3-A_4+2A_5)(\alpha '), \notag \\
	E_4(\alpha_2)=&\int_{0}^{\alpha_2}d\alpha_2'\int_{0}^{1-\alpha_2'}d\alpha_1(A_3-A_4)(\alpha '), \notag \\
	E_5(\alpha_2)=&\int_{0}^{\alpha_2}d\alpha_2'\int_{0}^{\alpha_2'}d\alpha_2''\int_{0}^{1-\alpha_2''}d\alpha_1(A_1-A_2+A_3 \notag \\& +A_4-A_5+A_6)(\alpha ''), \notag \\
	F_0(\alpha_2)=&\int_{0}^{1-\alpha_2}d\alpha_1 T_1(\alpha), \notag \\
	F_1(\alpha_2)=&\int_{0}^{\alpha_2}d\alpha_2'\int_{0}^{1-\alpha_2'}d\alpha_1(T_1+T_2-2T_3)(\alpha '), \notag \\
	F_2(\alpha_2)=&\int_{0}^{1-\alpha_2}d\alpha_1T_7(\alpha), \notag \\
	F_3(\alpha_2)=&\int_{0}^{\alpha_2}d\alpha_2'\int_{0}^{1-\alpha_2'}d\alpha_1(T_1-T_2-2T_7)(\alpha '), \notag \\
	F_4(\alpha_2)=&\int_{0}^{\alpha_2}d\alpha_2'\int_{0}^{1-alpha_2'}d\alpha_1(-T_1+T_5+2T_8)(\alpha '), \notag \\
	F_5(\alpha_2)=&\int_{0}^{\alpha_2}d\alpha_2'\int_{0}^{\alpha_2'}d\alpha_2''\int_{0}^{1-\alpha_2''}d\alpha_1(2T_2-2T_3-2T_4 \notag \\& +2T_5+2T_7+2T_8)(\alpha ''), \notag \\
	F_6(\alpha_2)=&\int_{0}^{\alpha_2}d\alpha_2'\int_{0}^{1-\alpha_2'}d\alpha_1(T_7-T_8)(\alpha '), \notag \\
	F_7(\alpha_2)=&\int_{0}^{\alpha_2}d\alpha_2'\int_{0}^{\alpha_2'}d\alpha_2''\int_{0}^{1-\alpha_2''}d\alpha_1(-T_1+T_2+T_5 \notag \\& -T_6+2T_7+2T_8)(\alpha ''),
	\end{align}
 where $\alpha, \alpha '$ and $\alpha ''$ are $(\alpha_1, \alpha_2, 1-\alpha_1-\alpha_2), (\alpha_1, \alpha_2', 1-\alpha_1-\alpha_2')$, and $(\alpha_1, \alpha_2'', 1-\alpha_1-\alpha_2'')$, respectively.The distribution amplitudes of $\Xi$ baryon are listed in Table \ref{tab1} with the increasing twists to twist six.  
 
  \section{Numerical analysis} \label{sec:III}

    In the numerical analysis, the parameters of baryons and leptons mass in the following analysis are adopted from PDG group. They are listed in Table \ref{tab2}.
    \begin{table}[h]
    	\caption{\label{tab2}Mass parameters of baryons and leptons.}
    	\begin{ruledtabular}
    	\begin{tabular}{cc} 
    		Parameters&PDG value \\ \hline 
    		$M_{\Xi_c^0}$&2.471 ~\rm{GeV} \\
    		$M_{\Xi_c^{*0}}$&2.7939 ~\rm{GeV} \\
    		$M_{\Xi_c^+}$&2.46771 ~\rm{GeV} \\
    		$M_{\Xi_c^{*+}}$&2.7919 ~\rm{GeV} \\
    		$M_{\Xi^-}$&1.3217 ~\rm{GeV} \\
    		$M_{\Xi^0}$&1.31486 ~\rm{GeV} \\
    		$m_c$&1.27 ~\rm{GeV} \\
    		$m_e$&0.51 ~\rm{MeV} \\
    		$m_\mu$&105.658 ~\rm{MeV} \\
    	\end{tabular}
        \end{ruledtabular}
    \end{table}
     
     The parameters in Table \ref{tab1} of the light-cone distribution amplitudes have been given in Ref.~\cite{Liu:2009uc} and listed in the following:
     \begin{align}
     \phi_3^0&=\phi_6^0=f_\Xi, \notag \\
     \psi_4^0&=\psi_5^0=\frac{1}{2}(f_\Xi-\lambda_1), \notag \\
     \phi_4^0&=\phi_5^0=\frac{1}{2}(f_\Xi+\lambda_1), \notag \\
     \phi_3^{'0}&=\phi_6^{'0}=-\xi_5^0=\frac{1}{6}(4\lambda_3-\lambda_2), \notag \\
     \phi_4^{'0}&=\xi_4^0=\frac{1}{6}(8\lambda_3-3\lambda_2), \notag \\
     \phi_5^{'0}&=-\xi_5^{'0}=\frac{1}{6}\lambda_2, \notag \\
     \xi_4^{'0}&=\frac{1}{6}(12\lambda_3-5\lambda_2),
     \end{align}
     
    where these parameters are
    \begin{align}
    f_\Xi&=(9.9\pm0.4)\times 10^{-3} ~\rm{GeV^2}, \notag \\
    \lambda_1&=-(2.8\pm0.1)\times 10^{-2}~\rm{GeV^2}, \notag \\
    \lambda_2&=(5.2\pm0.2)\times 10^{-2}~\rm{GeV^2}, \notag \\
    \lambda_3&=(1.7\pm0.1)\times 10^{-2}~\rm{GeV^2}.
    \end{align}
    
\begin{figure*}
 			\includegraphics[width=0.3\textwidth]{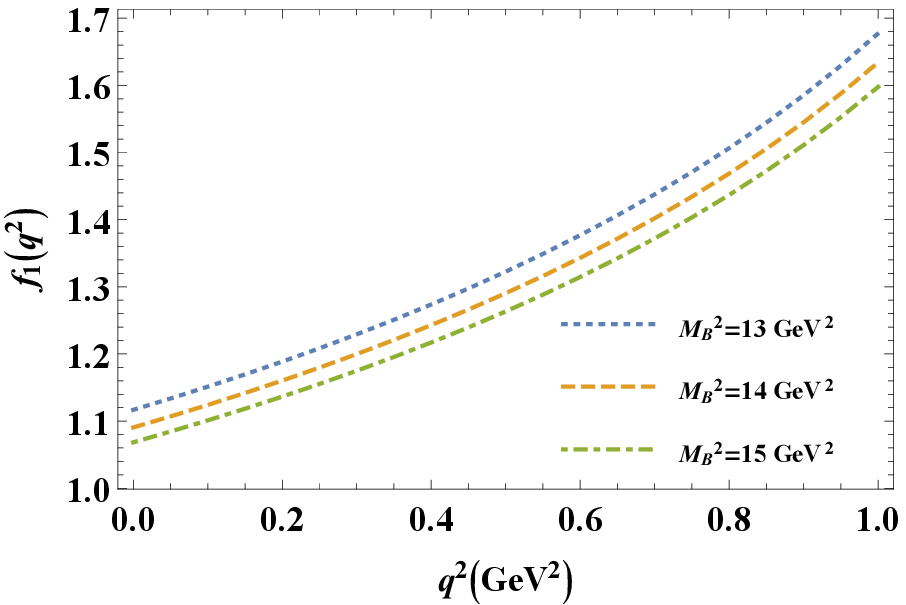}
 			\qquad
 			\includegraphics[width=0.3\textwidth]{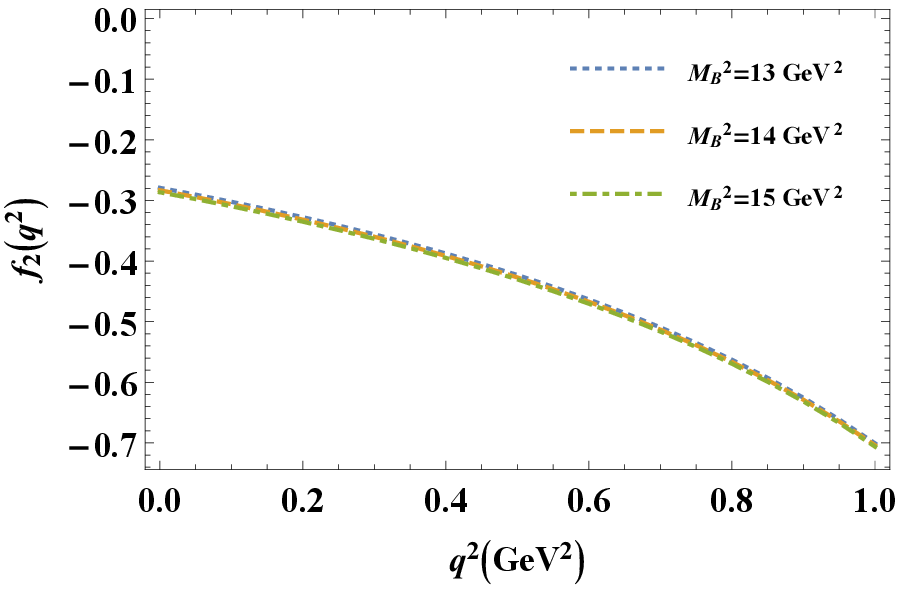}
 			\qquad
 			\includegraphics[width=0.3\textwidth]{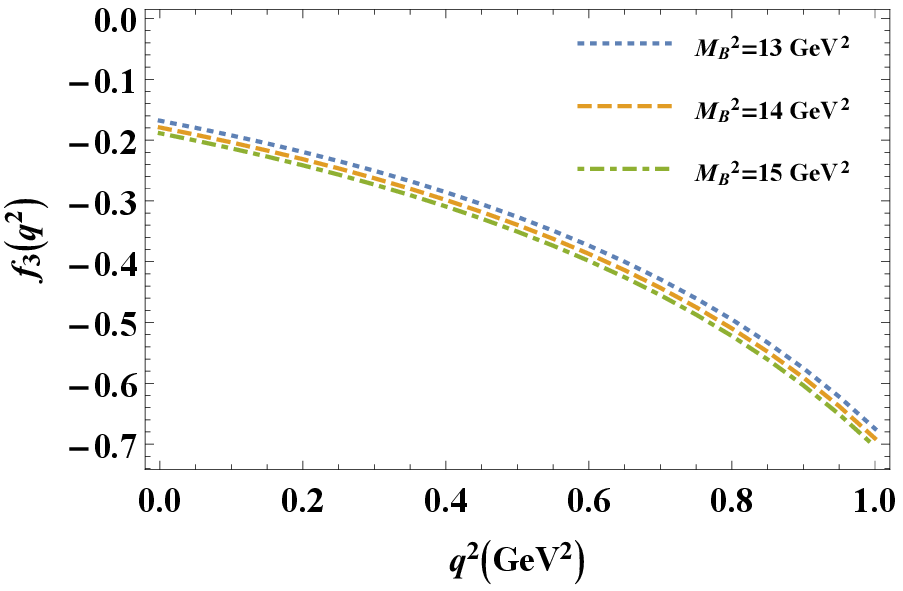}
 			\\
 			\includegraphics[width=0.3\textwidth]{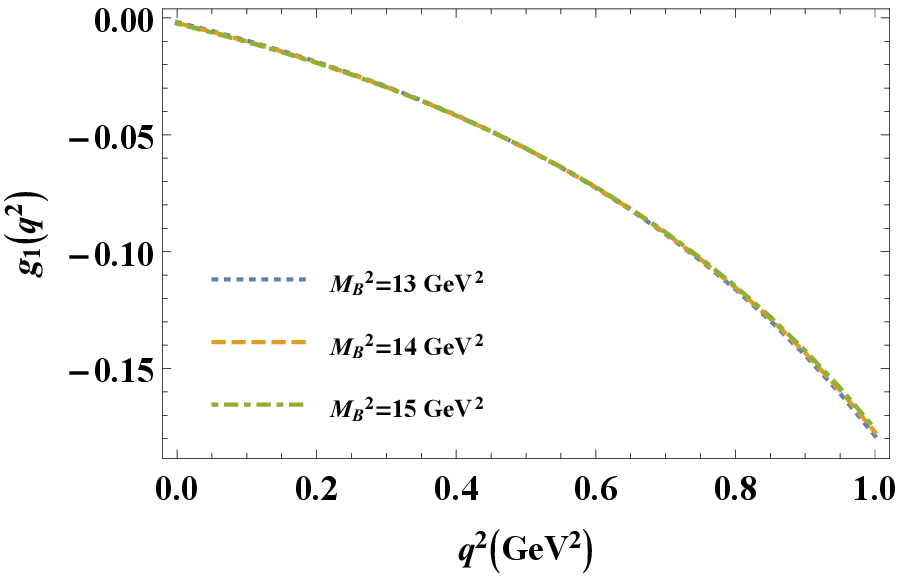}
 			\qquad
 			\includegraphics[width=0.3\textwidth]{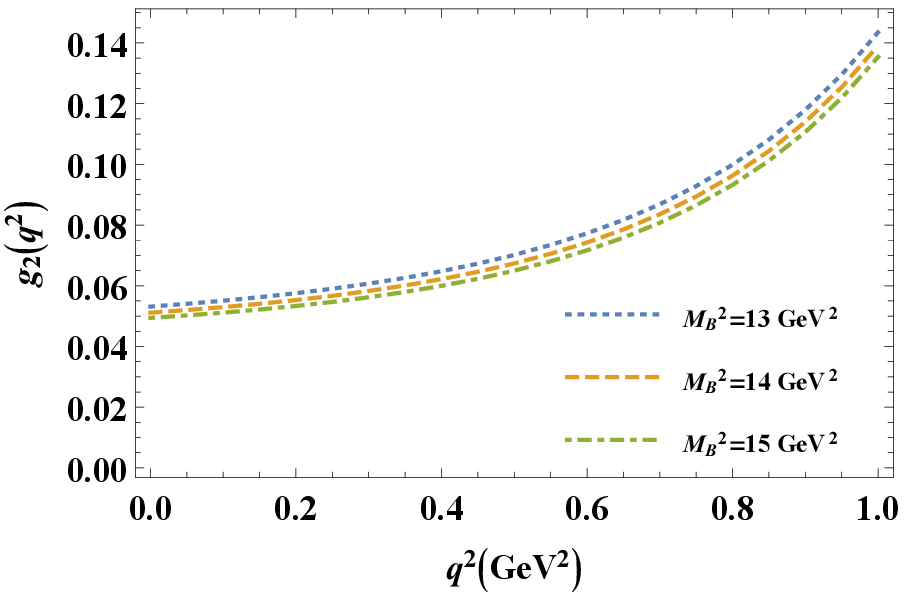}
 			\qquad
 			\includegraphics[width=0.3\textwidth]{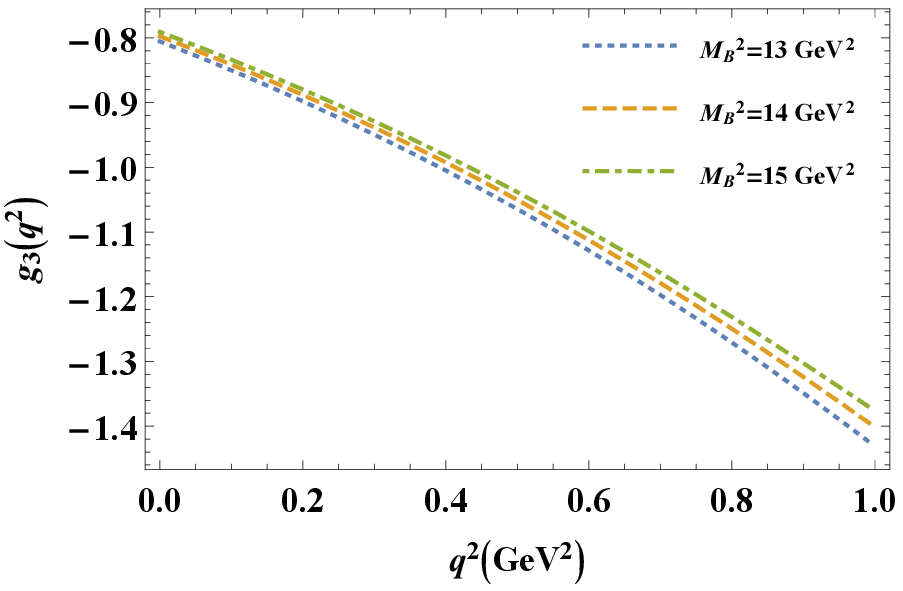}
 		\caption{\label{fig1}The form factors $f_i(i=1,2,3)$ and $g_i(i=1,2,3)$ of $\Xi_c^0 \to \Xi^-$ transition vary from zero-momentum transfer square to $q^2=1~{\rm{GeV}}^2$ with threshold $\sqrt{s_0}=(M_{\Xi_c}+0.45)~\rm{GeV}$}
 	\end{figure*}

    By using these parameters, the pictures of form factors $f_i(i=1,2,3)$ and $g_i(i=1,2,3)$ can be plotted by the expressions from Eq. (\ref{f1}) to (\ref{g3}). Because light-cone QCD sum rules are not suitable on the whole physical region, $m_\ell^2 \leq q^2 \leq (M_{\Xi_c}-M_\Xi)^2$. In this work, we adopt the light-cone QCD sum rules matching region $0~\rm{GeV^2} \leq q^2 \leq 1~\rm{GeV^2}$. For the purpose to suppress the higher twist contribution, the Borel transformations have been made on the hadronic level and theoretical level. In that case, an addition Borel parameter $M_B$ is introduced. The proper choice of Borel parameter will make the results reliable and stable. So, the choice of $M_B^2$ in this work makes $13~\rm{GeV^2} \leq M_B^2\leq 15~\rm{GeV^2}$. Another parameter we need is the threshold of heavy baryon $s_0$; it should be larger than the energy of $\Xi_c$ and $\Xi_c^*$ baryon, but smaller than the excited states mass of $\Xi_c$. In consideration of this, the setting value in this work is $0.4~\rm{GeV} \leq (\sqrt{s_0}-M_{\Xi_c}) \leq 0.5~\rm{GeV}$. With the basic parameters set, the pictures of $\Xi_c \to \Xi$ transition form factors in the light-cone QCD sum rules region are plotted in Fig \ref{fig1}.
    
     The momentum transfer square pictured in Fig. \ref{fig1} should be extrapolated on the whole physical region. For this purpose, the ``z-expansion'' fitting formula is used to achieve this aim \cite{Bourrely:2008za}. Keeping to the second order of ``z-series'', which is
      \begin{align}
      f_i(q^2)/g_i(q^2)=&\frac{1}{1-q^2/M_{D_s}^2}[a_0+a_1 z(q^2,t_0) \notag \\& +a_2 z(q^2,t_0)^2],
      \end{align}
     where $a_0, a_1$ and $a_2$ are fitting parameters and are listed with $f_i(0)/g_i(0)$ in Table \ref{table3}. $M_{D_s}=1.969$ $\rm{GeV^2}$ is the mass of $D_s$ meson. $t_0=(M_{\Xi_c}+M_\Xi)(\sqrt{M_{\Xi_c}}-\sqrt{M_\Xi})^2$, and
     	\begin{equation}
     	z(q^2,t_0)=\frac{\sqrt{t_+ -q^2}-\sqrt{t_+-t_0}}{\sqrt{t_+-q^2}+\sqrt{t_+-t_0}},
     	\end{equation}
     where $t_+=(M_{\Xi_c}+M_\Xi)$.	
     \begin{table*}
     		\caption{\label{table3}Form factors at zero momentum transfer square and fitting paramters $a_0, a_1$ and $a_2$ in the correspondance ``z-expansion" fitting fomulas at $\sqrt{s_0}=(M_{\Xi_c}+0.45)~\rm{GeV}$ and $M_B^2=14~\rm{GeV^2}$.}  
     	     	\begin{ruledtabular}
     		   	\begin{tabular}{cccccccccc} 
     		   	$f_i(q^2)$&$f_i(0)$&$a_0$&$a_1$&$a_2$&$g_i(q^2)$&$g_i(0)$&$a_0$&$a_1$&$a_2$ \\ \hline
     	    	$f_1(q^2)$&1.091&1.346&-4.048&17.924&$g_1(q^2)$&-0.002&-0.211&3.031&-9.870 \\
     		     $f_2(q^2)$&-0.279&-0.663&5.330&-16.315&$g_2(q^2)$&0.051&0.158&-2.320&12.607 \\
     		   	$f_3(q^2)$&-0.179&-0.724&8.149&-28.309&$g_3(q^2)$&-0.798&-1.127&2.731&4.386 \\ 
     		  	\end{tabular}
     	     \end{ruledtabular}
     \end{table*}
     
\begin{figure*}[!]
     	\includegraphics[width=0.4\textwidth]{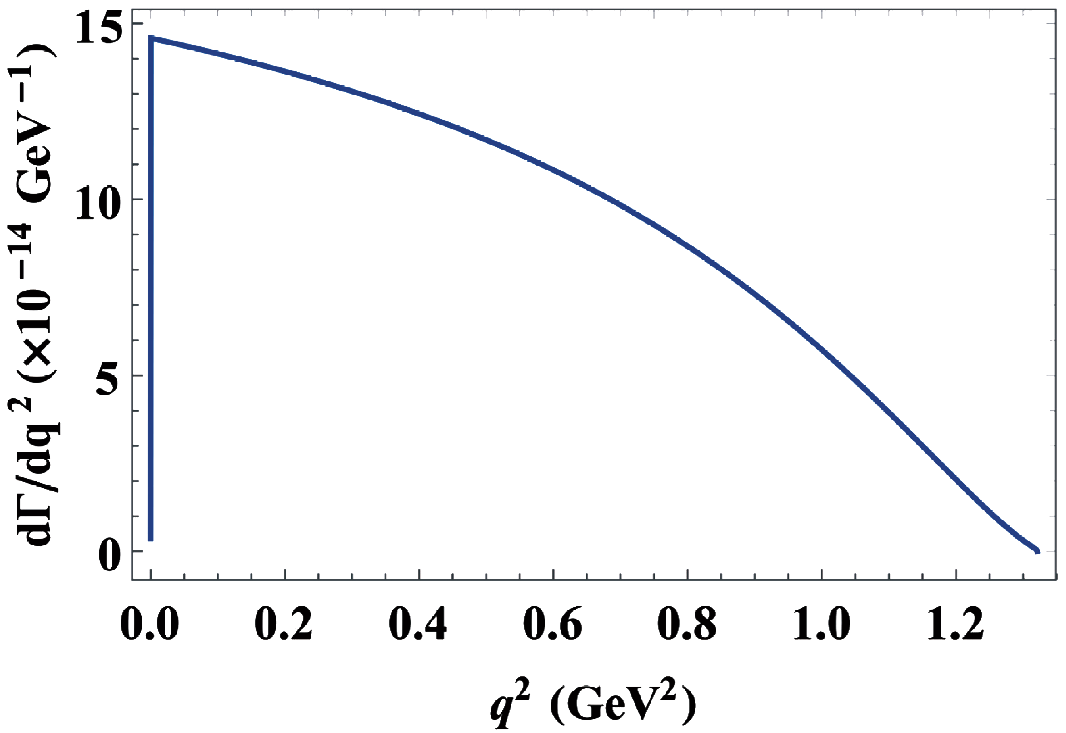}
     	\qquad
     	\includegraphics[width=0.4\textwidth]{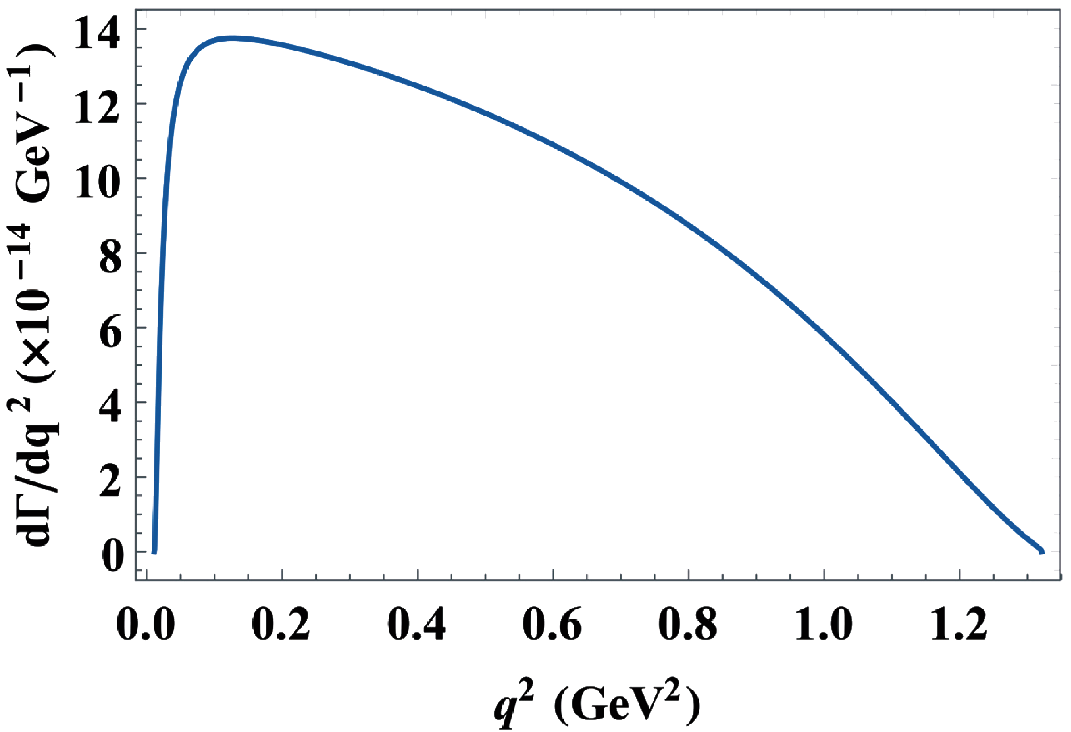}
       	\caption{\label{figure2}Differential decay width of $\Xi_c^0 \to \Xi^- e^+ \nu_e$ (left) and $\Xi_c^0 \to \Xi^- \mu^+ \nu_\mu$ (right) at $\sqrt{s_0}=(M_{\Xi_c}+0.45)~\rm{GeV}$ and $M_B^2=14~\rm{GeV^2}$.} 
    \end{figure*}

      In Fig. \ref{figure2} the differential decay widths of semileptonic decays of $\Xi_c^0$ are explicated. The pictures and results of $\Xi_c^+ \to \Xi^0 \ell^+\nu_\ell$ semileptonic decays can be obtained with the same steps. Absolute branching ratios of $\Xi_c \to \Xi \ell^+ \nu_\ell$ rely on the mean lifetime of $\Xi_c^0$ and $\Xi_c^+$, and the PDG average values give the suggestion that the two charm baryon lifetimes are $\tau_{\Xi_c^0}=(153\pm6)\times 10^{-15}~\rm{s}$ and $\tau_{\Xi_c^+}=(456\pm5)\times 10^{-15}~\rm{s}$.
     
      With the helicity amplitude representation of semileptonic decay widths, we have the following equation of differential decay width which can be written as two polarized decay width:
      \begin{gather}
      \frac{d\Gamma}{dq^2}=\frac{d\Gamma_L}{dq^2}+\frac{d\Gamma_T}{dq^2}
      \end{gather}
      and the total decay width is
      \begin{equation}
      \Gamma=\int_{m_l^2}^{(M_{\Xi_c}-M_\Xi)^2}dq^2\frac{d\Gamma}{dq^2},
      \end{equation}
      where
      \begin{align}
      \frac{d\Gamma_L}{dq^2}&=\frac{G_F^2|V_{cs}|^2 q^2 p(1-\hat{m}_l^2)^2}{384\pi^3 M_{\Xi_c}^2}[(2+\hat{m}_l^2)(|H_{\frac{1}{2},0}|^2 \notag \\& +|H_{-\frac{1}{2},0}|^2)+3\hat{m}_l^2(|H_{\frac{1}{2},t}|^2+|H_{-\frac{1}{2},t}|^2)],  \\
      \frac{d\Gamma_T}{dq^2}&=\frac{G_F^2|V_{cs}|^2 q^2 p(1-\hat{m}_l^2)^2(2+\hat{m}_l^2)}{384\pi^3 M_{\Xi_c}^2}(|H_{\frac{1}{2},1}|^2 \notag \\&+|H_{-\frac{1}{2},-1}|^2).
      \end{align}  
      
      In the above equations, $p=\sqrt{Q_+Q_-}/2M_{\Xi_c}, Q_\pm=(M_{\Xi_c}\pm M_\Xi)^2-q^2$ and $\hat{m}_l\equiv m_l/\sqrt{q^2}$, $m_l$ are the mass of leptons. $G_F$ is the fermi constant and $|V_{cs}|$ is the Cabibbo-Kobayashi-Maskawa (CKM) matrix element; they are $G_F=1.66\times 10^{-5}~\rm{GeV^{-2}}$ and $|V_{cs}|=0.987$, respectively~\cite{Zyla:2020zbs}. The related expressions of helicity amplitudes connected with form factors are given by
      \begin{align}
      H_{\frac{1}{2},0}^V=&-i\frac{\sqrt{Q_-}}{\sqrt{q^2}}[(M_{\Xi_c}+M_\Xi)f_1(q^2)-\frac{q^2}{M_{\Xi_c}}f_2(q^2)], \\
      H_{\frac{1}{2},1}^V=&i\sqrt{2Q_-}[-f_1(q^2)+\frac{M_{\Xi_c}+M_\Xi}{M_{\Xi_c}}f_2(q^2)],  \\
      H_{\frac{1}{2},t}^V=&-i\frac{\sqrt{Q_+}}{\sqrt{q^2}}[(M_{\Xi_c}-M_\Xi)f_1(q^2)+\frac{q^2}{M_{\Xi_c}}f_3(q^2)], \\
      H_{\frac{1}{2},0}^A=&-i\frac{\sqrt{Q_+}}{\sqrt{q^2}}[(M_{\Xi_c}-M_\Xi)g_1(q^2)+\frac{q^2}{M_{\Xi_c}}g_2(q^2)], \\
      H_{\frac{1}{2},1}^A=&i\sqrt{2Q_+}[-g_1(q^2)-\frac{M_{\Xi_c}-M_\Xi}{M_{\Xi_c}}g_2(q^2)], \\
      H_{\frac{1}{2},t}^A=&-i\frac{\sqrt{Q_-}}{\sqrt{q^2}}[(M_{\Xi_c}+M_\Xi)g_1(q^2)-\frac{q^2}{M_{\Xi_c}}g_3(q^2)].
      \end{align}
      The negative helicity amplitudes can be given by the positive helicity amplitudes as
      \begin{gather}
      H_{-\lambda,-\lambda_W}^V=H_{\lambda,\lambda_W}^V,   \notag \\     H_{-\lambda,-\lambda_W}^A=-H_{\lambda,\lambda_W}^A.
      \end{gather}
      $\lambda$ and $\lambda_W$ are the polarizations of the final $\Xi$ baryon and W-Boson, respectively.
    
     \begin{table*}[!]
     	\caption{\label{table4}Absolute branching ratios of semileptonic decay $\mathcal{B}(\Xi_c \to \Xi \ell^+ \nu_\ell) (\times 10^{-2})$.}
     	\begin{ruledtabular}
     		\begin{tabular}{ccccc} 
     			Works&$\mathcal{B}(\Xi_c^0 \to \Xi^-e^+ \nu_e)$&$\mathcal{B}(\Xi_c^0 \to \Xi^-\mu^+ \nu_\mu)$&$\mathcal{B}(\Xi_c^+ \to \Xi^0 e^+ \nu_e)$&$\mathcal{B}(\Xi_c^+ \to \Xi^0 \mu^+ \nu_\mu)$ \\ \hline
     			Exps&$(2.48\pm0.72)$\cite{ALICE:2021bli},$(1.31\pm0.38)$\cite{Belle:2021crz} &$(1.27\pm0.37)$\cite{Belle:2021crz}&-&- \\
     			This work&$2.81^{+0.17}_{-0.15}$&$2.72^{+0.17}_{-0.15}$&$8.43^{+0.52}_{-0.45}$&$8.16^{+0.50}_{-0.43}$ \\
     			Lattice&$(2.38\pm0.32)$\cite{Zhang:2021oja}&$(2.29\pm0.31)$\cite{Zhang:2021oja}&$(7.18\pm0.98)$\cite{Zhang:2021oja}&$(6.91\pm0.93)$\cite{Zhang:2021oja} \\
     			LCSR&\makecell[c]{2.03\cite{Liu:2009uc}, $(1.43^{+0.52}_{-0.57})$ \cite{Liu:2010bh}, \\ $(7.26\pm2.54)$\cite{Azizi:2011mw},$(1.85\pm0.56)$\cite{Aliev:2021wat}}&\makecell{$(7.15\pm2.50)$\cite{Azizi:2011mw},\\ $(1.79\pm0.54)$\cite{Aliev:2021wat}}&\makecell{6.05\cite{Liu:2009uc}, $(4.27^{+1.55}_{-1.72})$\cite{Liu:2010bh},\\ $(5.51\pm1.65)$\cite{Aliev:2021wat}}&$(5.53\pm1.61)$\cite{Aliev:2021wat} \\
     			SU(3)&\makecell[c]{$(4.87\pm1.74)$\cite{Geng:2018plk}, $(3.0\pm0.3$, \\ $2.4\pm0.3, 2.7\pm 0.2)$\cite{Geng:2019bfz}, \\ $(4.10\pm0.46)$\cite{He:2021qnc}}&$(3.98\pm0.57)$\cite{He:2021qnc}&\makecell[c]{$(3.38^{+2.19}_{-2.26})$\cite{Geng:2018plk}, $(11.9\pm1.3$, \\ $9.8\pm1.1, 10.7\pm0.9)$\cite{Geng:2019bfz}}&- \\
     			LFQM&\makecell[c]{1.35\cite{Zhao:2018zcb},$(1.72\pm0.35)$\cite{Ke:2021pxk}, \\ $(3.49\pm0.95)$\cite{Geng:2020gjh}}&$(3.34\pm0.94)$\cite{Geng:2020gjh}&\makecell[c]{5.39\cite{Zhao:2018zcb},$(5.20\pm1.02)$\cite{Ke:2021pxk}, \\ $(11.3\pm3.35)$\cite{Geng:2020gjh}}&- \\
     			RQM&2.38\cite{Faustov:2019ddj}&2.31\cite{Faustov:2019ddj}&9.40\cite{Faustov:2019ddj}&9.11\cite{Faustov:2019ddj} \\
     			QCDSR&$(3.4\pm0.7)$\cite{Zhao:2021sje}&-&$(10.2\pm2.2)$\cite{Zhao:2021sje}& \\  
     			PDG&$(1.8\pm1.2)$\cite{Zyla:2020zbs}&-&$(7\pm 4)$\cite{Zyla:2020zbs}&- \\
     		\end{tabular}
     	\end{ruledtabular}	
     \end{table*}
    
      With the $V-A$ current, the total helicity amplitudes are expressed as 
      \begin{gather}
      H_{\lambda,\lambda_W}=H_{\lambda,\lambda_W}^V-H_{\lambda,\lambda_W}^A.
      \end{gather}

     Substituting the numerical value of every parameters to the differential decay width with helicity formalism, one can obtain the decay width at $\sqrt{s_0}=(M_{\Xi_c}+0.45)~\rm{GeV}$ and $M_B^2=14~\rm{GeV^2}$ the decay widths $\Gamma(\Xi_c^0 \to \Xi^- e^+ \nu_e)=1.21\times 10^{-13}~\rm{GeV^2}$, and $\Gamma(\Xi_c^0 \to \Xi^- \mu^+ \nu_\mu)=1.17\times 10^{-13}~\rm{GeV^2}$. The charged charm baryon $\Xi_c^+$ decay widths $\Gamma(\Xi_c^+ \to \Xi^0 e^+ \nu_e)=1.22\times 10^{-13}~\rm{GeV^2}$, and $\Gamma(\Xi_c^+ \to \Xi^0 \mu^+ \nu_\mu)=1.18\times 10^{-13}~\rm{GeV^2}$. With these decay widths and the lifetimes of $\Xi_c^0$ and $\Xi_c^+$ baryons, the absolute branching ratios give $\mathcal{B}(\Xi_c^0 \to \Xi^- e^+ \nu_e)=(2.81^{+0.17}_{-0.15}) \%$, $\mathcal{B}(\Xi_c^0 \to \Xi^- \mu^+ \nu_\mu)=(2.72^{+0.17}_{-0.15}) \%$, $\mathcal{B}(\Xi_c^+ \to \Xi^0 e^+ \nu_e)=(8.43^{+0.52}_{-0.45}) \%$ and $\mathcal{B}(\Xi_c^+ \to \Xi^0 \mu^+ \nu_\mu)=(8.16^{+0.50}_{-0.43}) \%$. The errors come from the chosen range of threshold $s_0$ and Borel parameters $M_B$ are about six percent, so the other parameters' errors are all included. This gives the ratios $\mathcal{B}(\Xi_c^0 \to \Xi^- e^+ \nu_e)/\mathcal{B}(\Xi_c^0 \to \Xi^- \mu^+ \nu_\mu)=1.03$, and $\mathcal{B}(\Xi_c^+ \to \Xi^0 e^+ \nu_e)/\mathcal{B}(\Xi_c^+ \to \Xi^0 \mu^+ \nu_\mu)=1.03$. It's a good case to establish the lepton flavor universality.

     \section{Conclusion} \label{sec:IV}
      In conclusion, the semileptonic decays of $\Xi_c\to \Xi$ are investigated in the framework of light-cone QCD sum rules. The form factors of $\Xi_c \to \Xi$ weak decay are calculated by this method. Using these form factors and the helicity formalism differential decay width, the absolute branching ratios of semileptonic decay of $\Xi_c^0 \to \Xi^- \ell^+ \nu_\ell$ and $\Xi_c^+ \to \Xi^0 \ell^+ \nu_\ell$ are calculated and list in Table \ref{table4}. Our results are larger than what Belle reported but are consistent with ALICE's result, and are also in consistent with PDG averaged. The comparisons with other theoretical works such as lattice QCD calculation (Lattice), light-cone QCD sum rules (LCSR), SU(3) flavor symmetry (SU(3)), light-front quark model (LFQM), relativistic quark model (RQM), QCD sum rules (QCDSR) and experiments (Exps) are also listed in Table \ref{table4}. The ratios of positron final state and muon final state processes show that the lepton flavor universality is held and consistent with the experimental results.
  
     \begin{acknowledgments}
  
     This work was supported in part by the National Natural Science Foundation of China under Contract No. 11675263. H. H. D. thanks Prof. Wei Wang for useful discussion.
     
     \end{acknowledgments}
     
      \begin{widetext}
     \section*{Appendix: The detailed expressions of $\rho^i_\Gamma$} \label{appendix}
     \appendix
     \setcounter{equation}{0}
     \renewcommand\theequation{\arabic{equation}}
     The $\rho_\Gamma^i(i=1,2,3)$ in equation (\ref{sum rule})
      \begin{align}
      	\rho_{p_\nu}^1(\alpha_2,q^2)=&M_\Xi[2\alpha_2D_0(\alpha_2)+3D_1(\alpha_2)-3\alpha_2D_2(\alpha_2)]-2m_cF_0(\alpha_2), \\
      	\rho_{p_\nu}^2(\alpha_2,q^2)=&-M_\Xi\{2[\alpha_2^2M_\Xi^2-\alpha_2(M_\Xi^2-M_{\Xi_c}^2+q^2)]+q^2\}D_1(\alpha_2)+\alpha_2^2M_\Xi^3[D_3(\alpha_2)-2D_4(\alpha_2)] \notag  \\& +6\alpha_2M_\Xi^3D_5(\alpha_2)+M_\Xi^2m_c[2\alpha_2F_3(\alpha)-2\alpha_2F_4(\alpha_2)+F_5(\alpha_2)], \\
      	\rho_{p_\nu}^3(\alpha_2,q^2)=&-4\alpha_2M_\Xi^3[\alpha_2^2M_\Xi^2+q^2-\alpha_2(M_\Xi^2-M_{\Xi_c}^2+q^2)]D_5(\alpha_2)-2M_\Xi^2m_cq^2F_5(\alpha_2)+4\alpha_2^2M_\Xi^4m_cF_7(\alpha_2),  \\
      	\rho_{\gamma_\nu}^1(\alpha_2,q^2)=&-M_\Xi m_c[B_0(\alpha_2)+F_0(\alpha_2)-3F_2(\alpha_2)]-M_\Xi^2[D_1(\alpha_2)-\alpha_2D_2(\alpha_2)+\frac{1}{4}D_3(\alpha_2)]-[\alpha_2M_\Xi^2  \notag \\& -\frac{1}{2}(M_\Xi^2-M_{\Xi_c}^2+q^2)]D_0(\alpha_2),  \\
      	\rho_{\gamma_\nu}^2(\alpha_2,q^2)=&M_\Xi^3m_c[\alpha_2B_1(\alpha_2)-\alpha_2F_4(\alpha_2)+\frac{1}{2}F_5(\alpha_2)-3\alpha_2F_6(\alpha_2)-3F_7(\alpha_2)]+\alpha_2M_\Xi^2[\alpha_2M_\Xi^2 \notag \\& -\frac{1}{2}(M_\Xi^2-M_{\Xi_c}^2+q^2)]D_1(\alpha_2)-M_\Xi^2[\alpha_2^2M_\Xi^2-\alpha_2(M_\Xi^2-M_{\Xi_c}^2+q^2)+q^2](\frac{1}{2}D_3(\alpha_2) \notag \\& -\frac{1}{2}D_4(\alpha_2)]-\frac{3}{2}\alpha_2M_\Xi^4D_5(\alpha_2)+M_\Xi m_c[\alpha_2 M_\Xi^2-(M_\Xi^2-M_{\Xi_c}^2+q^2)]F_3(\alpha_2),  \\
      	\rho_{\gamma_\nu}^3(\alpha_2,q^2)=&2[\alpha_2^2M_\Xi^2-\alpha_2(M_\Xi^2-M_{\Xi_c}^2+q^2)+q^2][\alpha_2 M_\Xi^4 D_5(\alpha_2)+M_\Xi^3 m_cF_7(\alpha_2)], \\
      	\rho_{p_\nu\slashed{q}}^1(\alpha_2,q^2)=&-D_0(\alpha_2), \\
      	\rho_{p_\nu\slashed{q}}^2(\alpha_2,q^2)=&-\alpha_2M_\Xi^2[D_1(\alpha_2)+D_3(\alpha_2)-2D_4(\alpha_2)]-M_\Xi m_c[F_1(\alpha_2)+F_3(\alpha_2)], \\
      	\rho_{p_\nu\slashed{q}}^3(\alpha_2,q^2)=&2\alpha_2M_\Xi^3 m_c[F_5(\alpha_2)-2F_7(\alpha_2)], \\
      	\rho_{\gamma_\nu\slashed{q}}^1(\alpha_2,q^2)=&-M_\Xi D_2(\alpha_2), \\
      	\rho_{\gamma_\nu\slashed{q}}^2(\alpha_2,q^2)=&-M_\Xi^2 m_c[B_1(\alpha_2)-F_3(\alpha_2)-F_4(\alpha_2)-3F_6(\alpha_2)]-M_\Xi [ \alpha_2M_\Xi^2-\frac{1}{2}(M_\Xi^2-M_{\Xi_c}^2+q^2)]D_1(\alpha_2) \notag \\& +\frac{3}{2}M_\Xi^3 D_5(\alpha_2), \\
      	\rho_{\gamma_\nu\slashed{q}}^3(\alpha_2,q^2)=&-2[\alpha_2^2M_\Xi^2-\alpha_2(M_\Xi^2-M_{\Xi_c}^2+q^2)+q^2]M_\Xi^3 D_5(\alpha_2), \\
      	\rho_{q_\nu}^1(\alpha_2,q^2)=&-M_\Xi[D_0(\alpha_2)-3D_2(\alpha_2)], \\
      	\rho_{q_\nu}^2(\alpha_2,q^2)=&[\alpha_2 M_\Xi^3-M_\Xi(M_\Xi^2-M_{\Xi_c}^2+q^2)]D_1(\alpha_2)-M_\Xi^3[\alpha_2D_3(\alpha_2)-2\alpha_2D_4(\alpha_2)+6D_5(\alpha_2)] \notag \\& +M_\Xi^2 m_c[F_1(\alpha_2)-F_3(\alpha_2)+2F_4(\alpha_2)], \\
      	\rho_{q_\nu}^3(\alpha_2,q^2)=&4M_\Xi^3[\alpha_2^2M_\Xi^2-\alpha_2(M_\Xi^2-M_{\Xi_c}^2+q^2)+q^2]D_5(\alpha_2)-2M_\Xi^2m_c[\alpha_2M_\Xi^2-(M_\Xi^2-M_{\Xi_c}^2+q^2)]F_5(q^2) \notag \\& +4\alpha_2M_\Xi^4m_cF_7(\alpha_2), \\
      	\rho_{q_\nu\slashed{q}}^1(\alpha_2,q^2)=&0, \\
      	\rho_{q_\nu\slashed{q}}^2(\alpha_2,q^2)=&M_\Xi^2[D_1(\alpha_2)+D_3(\alpha_2)-2D_4(\alpha_2)], \\
      	\rho_{q_\nu\slashed{q}}^3(\alpha_2,q^2)=&2M_\Xi^3m_c[2F_7(\alpha_2)-F_5(\alpha_2)], \\
      	\rho_{p_\nu\gamma_5}^1(\alpha_2,q^2)=&M_\Xi[2\alpha_2E_0(\alpha_2)-3E_1(\alpha_2)+3\alpha_2E_2(\alpha_2)], \\
      	\rho_{p_\nu\gamma_5}^2(\alpha_2,q^2)=&M_\Xi\{q^2+2[\alpha_2^2M_\Xi^2-\alpha_2(M_\Xi^2+q^2-M_\Xi^2)]\}E_1(\alpha_2)+\alpha_2M_\Xi^3[\alpha_2E_3(\alpha_2)-2\alpha_2E_4(\alpha_2)-6E_5(\alpha_2)], \\
      	\rho_{p_\nu\gamma_5}^3(\alpha_2,q^2)=&4\alpha_2M_\Xi^3[\alpha_2^2M_\Xi^2+q^2-\alpha_2(M_\Xi^2-M_{\Xi_c}^2+q^2)]E_5(\alpha_2), \\
      	\rho_{\gamma_\nu\gamma_5}^1(\alpha_2,q^2)=&\alpha_2m_cC_0(\alpha_2)+[\alpha_2M_\Xi^2-(M_\Xi^2-M_{\Xi_c}^2+q^2)/2]E_0(\alpha_2)-M_\Xi^2[E_1(\alpha_2)-\alpha_2E_2(\alpha_2)-\frac{1}{4}E_3(\alpha_2)], \\
      	\rho_{\gamma_\nu\gamma_5}^2(\alpha_2,q^2)=&\alpha_2M_\Xi^2\{M_\Xi m_cC_1(\alpha_2)+[\alpha_2M_\Xi^2-(M_\Xi^2-M_{\Xi_c}^2+q^2)/2]E_1(\alpha_2)\}+\frac{1}{2}M_\Xi^2[\alpha_2^2M_\Xi^2+q^2 \notag \\& -\alpha_2(M_\Xi^2+q^2-M_{\Xi_c}^2)](E_3(\alpha_2)-E_4(\alpha_2))-\frac{3}{2}\alpha_2M_\Xi^4E_5(\alpha_2), \\
      	\rho_{\gamma_\nu\gamma_5}^3(\alpha_2,q^2)=&2\alpha_2M_\Xi^4[\alpha_2^2M_\Xi^2+q^2-\alpha_2(M_\Xi^2+q^2-M_{\Xi_c}^2)]E_5(\alpha_2), \\
      	\rho_{p_\nu\slashed{q}\gamma_5}^1(\alpha_2,q^2)=&E_0(\alpha), \\
      	\rho_{p_\nu\slashed{q}\gamma_5}^2(\alpha_2,q^2)=&-\alpha_2M_\Xi^2[E_1(\alpha_2)-E_3(\alpha_2)+2E_4(\alpha_2)], \\
      	\rho_{p_\nu\slashed{q}\gamma_5}^3(\alpha_2,q^2)=&0, \\
      	\rho_{\gamma_\nu \slashed{q}\gamma_5}^1(\alpha_2,q^2)=&M_\Xi E_2(\alpha_2), \\
      	\rho_{\gamma_\nu \slashed{q}\gamma_5}^2(\alpha_2,q^2)=&M_\Xi^2 m_c C_1(\alpha_2)+M_\Xi[\alpha_2M_\Xi^2-(M_\Xi^2+q^2-M_{\Xi_c}^2)/2]E_1(\alpha_2)-\frac{3}{2}M_\Xi^3E_5(\alpha_2), \\
      	\rho_{\gamma_\nu \slashed{q}\gamma_5}^3(\alpha_2,q^2)=&2M_\Xi^3[\alpha_2^2M_\Xi^2+q^2-\alpha_2(M_\Xi^2+q^2-M_\Xi^2)]E_5(\alpha_2), \\
      	\rho_{q_\nu\gamma_5}^1(\alpha_2,q^2)=&-M_\Xi[E_0(\alpha_2)+3E_2(\alpha_2)], \\
      	\rho_{q_\nu\gamma_5}^2(\alpha_2,q^2)=&-M_\Xi[\alpha_2M_\Xi^2-(M_\Xi^2+q^2-M_{\Xi_c}^2)]E_1(\alpha_2)-M_\Xi^3[\alpha_2E_3(\alpha_2)-2\alpha_2E_4(\alpha_2)-6E_5(\alpha_2)], \\
      	\rho_{q_\nu\gamma_5}^3(\alpha_2,q^2)=&-4M_\Xi^3[\alpha_2^2M_\Xi^2+q^2-\alpha_2(M_\Xi^2+q^2-M_{\Xi_c}^2)]E_5(\alpha_2), \\
      	\rho_{q_\nu\slashed{q}\gamma_5}^1(\alpha_2,q^2)=&0, \\
      	\rho_{q_\nu\slashed{q}\gamma_5}^2(\alpha_2,q^2)=&M_\Xi^2[E_1(\alpha_2)-E_3(\alpha_3)+2E_4(\alpha_2)], \\
      	\rho_{q_\nu\slashed{q}\gamma_5}^3(\alpha_2,q^2)=&0.
      \end{align} 
      \end{widetext}
 
    \nocite{*}
    \bibliography{charm}
    
\end{document}